\documentclass[
    reprint,
    superscriptaddress,
    amsmath,
    amssymb,
    longbibliography,
    aps,
    pra,
]{revtex4-2}

\usepackage{graphicx}
\usepackage{braket}
\usepackage{amsmath,amsfonts,amsthm,bm}
\usepackage{hyperref}
\hypersetup{
    colorlinks=true,
    linkcolor=blue,
    citecolor=cyan,
    urlcolor=cyan,
}
\usepackage{multirow}
\normalsize 
\usepackage{array} 
\usepackage{booktabs} 
\usepackage{mathrsfs}
\setcitestyle{authoryear,round}

\usepackage{acronym}
\newacro{LHC}{Large Hadron Collider}
\newacro{HEP}{High Energy Physics}
\newacro{QML}{Quantum Machine Learning}
\newacro{ML}{Machine Learning}
\newacro{SM}{Standard Model}
\newacro{BSM}{Beyond Standard Model}
\newacro{MC}{Monte Carlo}
\newacro{AD}{Anomaly Detection}
\newacro{AE}{autoencoder}
\newacro{VAE}{variational autoencoder}
\newacro{QSVM}{Quantum Support Vector Machine}
\newacro{SVM}{Support Vector Machine}
\newacro{TPR}{True Positive Rate}
\newacro{FPR}{False Positive Rate}
\newacro{PCA}{Principal Component Analysis}
\newacro{VQC}{Variational Quantum Circuit}
\newacro{MSE}{Mean Squared Error}
\newacro{ROC}{Receiver Operating Curve}
\newacro{AUC}{Area Under Curve}
\newacro{GQC}{Guided Quantum Compression}
\newacro{SE}{Spectral Embedding}
\newacro{PCA}{Principal Component Analysis}
\newacro{ICA}{Independent Component Analysis}
\newacro{LLE}{Locally Linear Embedding}
\newacro{NMF}{Non-Negative Matrix Factorisation}
\newacro{RBM}{Restricted Boltzmann Machine}
\newacro{KL}{Kullback-Leibler}
\newacro{BCE}{binary cross entropy}
\newacro{AUC}{Area Under Curve}

\newcommand{\tth}{$t\bar{t}H(b\bar{b})$ }

\makeatletter
\renewcommand{\fnum@figure}{\textbf{Fig.} \thefigure}
\makeatother
\makeatletter
\renewcommand{\@caption@fignum@sep}{\space}
\makeatother

\begin{document}
\title{Learning Reduced Representations for Quantum Classifiers}

\author{Patrick Odagiu}
\email{podagiu@ethz.ch}
\affiliation{Institute for Particle Physics and Astrophysics, ETH Zurich, 8093 Zurich, Switzerland}
\author{Vasilis Belis}
\affiliation{Institute for Particle Physics and Astrophysics, ETH Zurich, 8093 Zurich, Switzerland}
\author{Lennart Schulze}
\affiliation{Columbia University, 10027 New York, NY, USA.}
\author{Panagiotis Barkoutsos}
\affiliation{IBM Quantum, IBM Research – Zurich, 8803 R\"uschlikon, Switzerland}
\author{Michele Grossi}
\affiliation{European Organization for Nuclear Research (CERN), CH-1211 Geneva, Switzerland}
\author{Florentin Reiter}
\affiliation{Institute for Quantum Electronics, ETH Zurich, 8093 Zurich, Switzerland}
\affiliation{Fraunhofer Institute for Applied Solid State Physics IAF, Tullastr. 72, 
79108 Freiburg, Germany}
\author{G\"unther Dissertori}
\affiliation{Institute for Particle Physics and Astrophysics, ETH Zurich, 8093 Zurich, Switzerland}
\author{Ivano Tavernelli}
\affiliation{IBM Quantum, IBM Research – Zurich, 8803 R\"uschlikon, Switzerland}
\author{Sofia Vallecorsa}
\affiliation{European Organization for Nuclear Research (CERN), CH-1211 Geneva, Switzerland}

\date{\today}

\begin{abstract}

Data sets that are specified by a large number of features are currently outside the area of applicability for quantum machine learning algorithms.
An immediate solution to this impasse is the application of dimensionality reduction methods before passing the data to the quantum algorithm.
We investigate six conventional feature extraction algorithms and five autoencoder-based dimensionality reduction models to a particle physics data set with 67 features.
The reduced representations generated by these models are then used to train a quantum support vector machine for solving a binary classification problem: whether a Higgs boson is produced in proton collisions at the LHC.
We~\mbox{show} that the autoencoder methods learn a better lower-dimensional representation of the data, with the method we design, the Sinkclass autoencoder, performing 40\% better than the baseline.
The methods developed here open up the applicability of quantum machine learning to a larger array of data sets. 
Moreover, we provide a recipe for effective dimensionality reduction in this context.\\

\end{abstract}

\maketitle
\section{Introduction}\label{sec:intro}

A large amount of relevant \ac{ML} data sets contain high-dimensional data~\citep{kaggle}.
Meanwhile, the number of features that specifies these data is much larger than what a currently available quantum computer can process: the quantum circuits offered by state of the art machines do not have sufficient depth and do not posses a large enough number of qubits.
\mbox{However}, a vast class of \ac{ML} data pertaining to biology, chemistry, and physics, remain inaccessible for \ac{QML} algorithms, which surpass classical \ac{ML} in a variety of specific tasks~\citep{Rebentrost2014, Liu2021rigorous, muser2023, Huang2021, huangQA2022, pirnay_super-polynomial_2022, Gyurik:2023quj}. 
Therefore, dimensionality reduction is often needed to exploit~\ac{QML} on interesting, high-dimensional data sets, especially in scientific data and at particle accelerators, where QML might make the biggest difference. 

Dimensionality reduction~\citep{zebari2020comprehensive} has been employed as a preprocessing step before passing the data to a \ac{QML}~model. 
Existing studies use different approaches to reducing the dimensionality of their data: manual feature selection based on specialist knowledge \citep{terashi2021, Belis:2021zqi}, linear feature extraction such as \ac{PCA} \mbox{\citep{wu2021_kernel, wu2021_vqc, Schuhmacher23}}, or more modern approaches involving dimensionality reduction through deep learning, such as using basic autoencoder architectures~\mbox{\citep{Belis:2021zqi, Belis:2023atb, ballard1987modular}}.

However, most studies regard dimensionality reduction merely as a preprocessing step, often assigning it less importance than the optimisation of the \ac{QML}~model. 
\mbox{Consequently}, the heuristics for optimising dimensionality reduction algorithms remain underexplored in the specific context of a given \ac{QML} model applied to data.
We systematically study and benchmark numerous conventional dimensionality reduction techniques that have been frequently employed in the literature, along with several deep learning-based autoencoder architectures. 
Furthermore, we introduce an autoencoder-based feature reduction algorithm, named \textit{Sinkclass}, which consistently outperforms all the other techniques we study herein. 
This algorithm broadens the applicability of \ac{QML} classifiers to a wider range of data sets without compromising classification performance.

A hybrid quantum-classical algorithm that uses dimensionality reduction is introduced in~\citet{Belis:2024ctj}.
In~contrast with the Guided Quantum Compression introduced therein, the methods investigated in this work allow for a free choice of \ac{QML} classifier algorithms, rather than being restricted to a variational quantum circuit. 
This is a significant advantage, since variationally trained quantum circuits sometimes exhibit trainability issues in gradient-based learning~\citep{bp_review24}, and typically require a large number of samples to estimate gradients at each parameter update.
Hence, we study six classical feature extraction algorithms and five different autoencoder models, including our new Sinkclass model.
A quantum support vector machine (QSVM) is chosen as the architecture to train on the output of these methods.
Thus, we show which types of dimensionality reduction induce the best classification performance for our QSVM.
\newpage


The studied dimensionality reduction methods are applied to a \ac{HEP} data set that defines a binary classification problem.
This data set is chosen for two main reasons: it is high-dimensional with 67 features per event and these features are measured quantum observables that describe highly energetic proton-proton collisions at the \ac{LHC}. 
Moreover, our features can be calculated from fundamental principles in the framework of quantum field theory.
Therefore, the quantum nature of the data generation process motivates the application of \ac{QML} \citep{Gyurik:2023quj,Bowles:2024fvp,Guan_2021} and its dimensionality imposes the need for reduction.

This work represents one of the first systematic studies of dimensionality reduction applied to a real-world data set, especially relevant for \ac{QML} application, and with generalization potential towards other high-dimensional data sets as well~\citep{kaggle}.
Additionally, our methods constitute new ways of effectively exploiting the data measured at the \ac{LHC}~\citep{Albertsson:2018maf} and provide another example of the synergy between conventional \ac{ML} and \ac{QML} \citep{Biamonte2017, Schuld_2018_book, Benedetti2019, Havlicek2019, Schuld_2019, circuit_centric_Schuld2020}.



\section{The Data}\label{sec:data}

The top quark is the heaviest particle described within the \ac{SM} of particle physics.
Hence, the interaction between the top quark and the Higgs boson is studied to understand how mass is generated~in~the~\ac{SM};
additionally, this interaction is also studied to search for new phenomena that cannot be predicted using~the~\ac{SM}.
For details on why the considered top-Higgs interaction is interesting, see \citet{Bezrukov:2014ina}.

The interaction strength between the Higgs boson and the top quark is called the coupling of these particles.
This coupling can be measured directly via Higgs boson production in association with a top quark and a top anti-quark, $t\bar t$ \citep{reissel_thesis}.
The Higgs boson, once produced by the top quark interaction with its anti-matter counterpart, can decay in a number of ways.
The~most likely decay is to a bottom quark anti-quark pair, $H \to b\bar b$.
Moreover, there are processes in the \ac{SM} that produce similar signatures in our detectors as~the~studied~\mbox{process}.
To~mitigate confounding with these background processes, we consider top quark anti-quark interactions where the Higgs boson is produced \textit{and} one of the top quarks decays into a charged lepton.

In summary, we consider the \tth process in the semi-leptonic decay channel.
The data defining this process and its most likely background, i.e.,  $t\bar{t}\rightarrow g \rightarrow b\bar{b}$, where the Higgs boson is replaced by a gluon, are produced by simulating proton collisions at the \ac{LHC} with centre-of-mass energy $\sqrt{s}=13$\,TeV~\citep{tthbb_data}. 
Each sample in our data corresponds to a simulated collision event; see \citet{reissel_thesis} for details.

In the context of this study, we choose to benchmark our methods on the \tth data set because its features are quantum observables that can be calculated from fundamental principles using the \ac{SM}. 
Moreover, each sample in this data contains 67 features, which is representative of popular high-dimensional data~\citep{kaggle}.
Lastly, using the same data set as \citet{Belis:2021zqi} allows us to compare the methods developed here with their baseline.

The task on the \tth data set is to identify collision events where the Higgs boson is produced. 
This objective can be formulated as a binary classification problem between \textit{signal} events and \textit{background} events. 
In the class of events referred to as signal, a Higgs boson is produced in the proton collision event. In background events no Higgs boson production occurs, but rather a gluon is produced. 

The features of each sample are physical \mbox{observables}, recorded by the Compact Muon Solenoid~\citep{CMS:2008xjf} at~the~\ac{LHC}. 
These physical observables are quantities that describe the quantum mechanical processes of particle production, interaction, and decay. 
Each data feature follows a probability distribution which can be computed via a combination of analytical and numerical methods.
This computation is done in the context of quantum field theory, specifically the \ac{SM} of particle physics, by employing numerical \ac{MC} simulations~\citep{Nason:2004rx, Frixione:2007vw, Alioli:2010xd, Sjostrand:2007gs, deFavereau:2013fsa}.
The $t\bar{t}H(b\bar{b})$ data set is represented by the following features~\citep{tthbb_data}:
\begin{enumerate}
    \item Jet features: ($p_\mathrm{T}$, $\eta$, $\phi$, $E$, b-tag, $p_\mathrm{x}$, $p_\mathrm{y}$, $p_\mathrm{z}$). 
    \item Lepton features: ($p_\mathrm{T}$, $\eta$, $\phi$, $E$, $p_\mathrm{x}$, $p_\mathrm{y}$, $p_\mathrm{z}$).
    \item Missing energy features (MET): ($\phi$, $p_\mathrm{T}$, $p_\mathrm{x}$, $p_\mathrm{y}$).
\end{enumerate}
Lepton here refers to electrons and muons, but not taus. 
In~the following, we define each of the quantities introduced above, the concept of a jet, and missing energy.

The direction of the initial proton beam is taken to be along the $z$ axis of the experiment. 
Hence, the transverse momentum $p_T$ and transverse mass $m_T$ of a particle are
\begin{equation}
p_{T}=\sqrt{p_x^2+p_y^2}=p\sin\theta\quad\text{and}\quad m_T\equiv\sqrt{p_T^2+m^2},
\end{equation}
where $p_{x/y}$ is the momentum along the $x$/$y$-axis, $p$ is the magnitude of the momentum, and $\theta$ is the angle measured from the $z$-axis, conventionally called the polar angle.  
The four-momentum of a particle can be described in three equivalent representations:
\begin{equation}
\label{eq:four_mom}
p^{\mu} = 
\begin{pmatrix}
E\\p_x \\ p_y \\p_z
\end{pmatrix}
= \begin{pmatrix}
E\\ p\sin{\theta}\cos{\phi} \\ p\sin{\theta}\sin{\phi} \\ p\cos{\theta}
\end{pmatrix}
=\begin{pmatrix} 
m_T\cosh{y} \\ p_T\cos{\phi} \\ p_T\sin{\phi} \\ m_T\sinh{y}
\end{pmatrix}
\end{equation}
where Greek indices enumerate the spacetime coordinates $\mu=\{0,1,2,3\}$, $\phi$ is the azimuthal angle, $y$ is the rapidity, and $E$ the energy of the particle. 
The rapidity $y$ of a particle boosted along the $z$ direction is defined as
\begin{equation}
\label{eq:rapidity}
y=\text{arctanh}\left(\frac{p_z}{E}\right)=\frac{1}{2}\ln\left(\frac{E+p_z}{E-p_z}\right).
\end{equation}
Furthermore, the pseudo-rapidity is defined as the limit of $y$ as the particle approaches masslessness
\begin{equation}
\eta=\lim_{m\to0}y=-\ln\tan\left(\frac{\theta}{2}\right).
\label{eq:pseudo_rapidity}
\end{equation}

The coordinate system $(p_T,\eta,\phi)$ is often used at hadron colliders, such as the LHC, since it displays convenient properties. 
Specifically, $p_T$ and $\phi$ are relativistically invariant quantities under Lorentz boosts along the $z$-axis, and $\eta$ is additive under Lorentz \mbox{transformations}. 
Furthermore, particles that interact weakly with matter cannot be detected directly using the \ac{LHC} detectors: they pass through the apparatus leaving missing transverse momentum and energy. 
Thus, the amount of missing transverse energy is an important feature of an event.

Finally, the $b$-tag is the only feature in this data that is not a quantum observable. 
As discussed in Sec.~I, quantum observables represent particle features that are measured by detectors and are computed from first principles using the quantum field theory framework of the~SM.
The~$b$-tag feature characterises the so called jets, which are a collection of collimated particles interacting with the detector in a constrained~area.
Namely, the b-tag feature of a jet is a binary variable that specifies whether the jet originates from the hadronisation of bottom quarks. 
\mbox{Dedicated} algorithms \citep{Cacciari:2008gp} construct the four momentum vector of a jet using the information from the detector; further algorithms are used to identify their~\mbox{origin}.
Hence, $b$-tag is not a quantum observable.
For~more~details on the $b$-tag see \citet{Belis:2024ctj}.

Before processing the data with either \ac{ML} or \ac{QML} classification algorithms, the features are filtered using some physically motivated criteria that take into account the geometric acceptance of the detector and the goal of background suppression.  
Specifically, we require: \mbox{$p_\mathrm{T} > 30$}\,GeV and $|\eta|<2.1$ for electrons, $p_\mathrm{T} > 26$\,GeV and $|\eta|<2.1$ for muons, $p_\mathrm{T} > 30$\,GeV and $|\eta|<2.4$ for jets, and the isolation of the leptons with respect to jets to be greater than $0.1$~\citep{Moortgat:1756841}.
The lepton isolation factor specifies how far the lepton is isolated with respect to any other object in the detector; for more details on how this is computed see \cite{CMS:2017yfk}.
Furthermore, we require that there must be at least four reconstructed jets per event, at least two $b$-tagged jets, and exactly one lepton~\citep{Belis:2021zqi}. 
Lastly, only the first seven highest transverse momentum jets are kept, allowing for one extra jet to include final state radiation. 
If there are more than seven reconstructed jets per data sample, i.e., collision event, zero padding is applied.
In conclusion, each sample in the data contains 67 features 
\begin{equation}
D = \underbrace{7\times 8}_\text{jets} + \underbrace{1\times 7}_\text{lepton} + \underbrace{1\times 4}_\text{MET} = 67.
\end{equation}


\section{Dimensionality Reduction}
\label{sec:dimred}

While the full $67$ features of the analysed data set could be given to a classical \ac{ML} model, the same is not true for models implemented on current quantum computers. 
In \ac{QML} algorithms, the number of available qubits defines the dimensionality of the input feature space, given an encoding strategy. 
With the low number of qubits currently available in present quantum computers, \ac{QML} algorithms can only process low-dimensional data sets. 
Hence, directly processing inherently high-dimensional data, such as images or \ac{HEP} data~\citep{Belis:2023atb}, with \ac{QML} models is a challenge.
To adapt to these limitations, previous studies of QML in \ac{HEP} either use a conventional feature reduction method~\citep{Wu:2020cye, Schuhmacher23} or select only a subset of the available features~\mbox{\citep{Terashi:2020wfi, Belis:2021zqi}}. 

\subsection{General Principles}

Dimensionality reduction refers to the set of methods used for mapping high-dimensional data into a lower-dimensional space such that discriminative properties are retained between samples in that space. 
When used for downstream ML methods, it results in performance improvements for the respective 
\ac{ML} task~\citep{Maaten2008DimensionalityRA, ray2021various, AEreview}.
There exist two primary approaches to dimensionality reduction achieved through \ac{ML} methods, which are described in the rest of this chapter. 

First, feature selection maintains interpretability by determining a subset of the most informative features from the original data.
Thus, this approach relies on prior knowledge regarding the importance of the data features in relation to the undertaken task, or on a systematic brute-force procedure to determine this \mbox{importance}.  
Conversely, feature extraction constructs new, lower-dimensional features that ideally capture the most significant information from the original data: interpretability is diminished, but this procedure can be automated without using domain knowledge or making prior assumptions about the data. 
In addition, feature extraction can make use of all input features and thus promises higher performance than feature selection.
Feature extraction is further subdivided into linear and nonlinear techniques. 

Linear feature extraction constructs new features from linear combinations of the original features.
Thus, they are relatively easy to implement, computationally efficient, and maintain a certain degree of interpretability.
The most popular such technique is \ac{PCA}~\citep{osti_15002155}, although other methods are also considered in this work.
Linear feature extraction methods only capture global correlations and this often results in information loss.
This is especially the case when processing data with non-linear relations between features, which applies to currently popular \ac{ML} data sets and particularly to the \tth data set.

Alternatively, nonlinear feature extraction applies nonlinear transformations to the data, thus generating a new set of features that better capture the complex correlations within the original data. 
Manifold techniques are a type of nonlinear feature extraction that assumes higher-dimensional data exists on a lower-dimensional manifold so that it can be embedded into that lower-dimensional space without losing essential information. 
This is achieved by attempting to preserve the local structure of the data; thus, these techniques are not guaranteed to capture global structure \citep{meilă2023manifoldlearningwhathow}.

The newest approach towards non-linear feature extraction is the usage of deep learning. 
This type of \ac{ML} models learn to construct a lower-dimensional representation of the input data that minimises a certain criterion, usually inversely proportional to the information content within the lower-dimensional data \mbox{representation}.
\mbox{\ac{ML}-based} feature extraction has three popular~\mbox{models}: convolutional networks, recurrent networks, and autoencoders (AEs) \citep{AEreview, bank2023autoencoders}.

This study focuses on autoencoders.
AEs perform automatic feature extraction without relying on a priori knowledge about the data, in contrast with the other two types of mentioned neural networks.  
Moreover, the AE has been used for tasks outside dimensionality reduction as well, e.g., anomaly detection \citep{Govorkova_2022, rev_ML_ad2024} and denoising \citep{vincent2008extracting}.
All these applications rely on the same principle: the lower-dimensional representation of the data produced by an AE is, in the ideal case, the most essential and informative for the considered task.


Although AEs present a promising approach to dimensionality reduction, they also exhibit disadvantages compared with more conventional methods, c.f., Sec.~\ref{sec:conventional}. 
Every AE architecture is sensitive to a relatively large array of hyperparameters such as its number of layers, the employed learning rate, and so on.
Currently, there exist methods of finding approximately optimal values for these hyperparameters, but they are usually computationally expensive and are not guaranteed to give the best performance~\citep{bank2023autoencoders, AEreview}.
Therefore, AEs generally lack robustness when compared with conventional dimensionality reduction paradigms.
They also require a large amount of data for training. 
Furthermore, AEs do not preserve the spatial or temporal locality of the data set unless explicitly enforced.
This~renders them impractical for applications in image segmentation tasks or sequential data tasks, respectively. 
In addition, AEs have a tendency to focus on lower-order data features and may struggle to represent highly complex correlations in the data. 
Therefore, if a data set is known to have intricate global dependencies, then a basic AE architecture is less favourable for effectively performing the dimensionality reduction.
Finally, the AE training is computationally more expensive than the fitting process of most classical feature extraction methods. 
\mbox{Nonetheless}, the particle physics data set analysed herein presents good synergy with AE methods.

\subsection{Conventional Methods}
\label{sec:conventional}

We study six classical feature extraction algorithms: PCA, independent component analysis \citep{hyvarinen1999independent}, locally linear embedding~\citep{donoho2003hessian}, spectral embedding \citep{ng2001spectral}, non-negative matrix factorisation \citep{6165290}, and restricted boltzmann machines \citep{tieleman2008training}. 
They are described in more detail in App.\,\ref{sec:classicalfe}.
We selected these methods since they are popular in the ML literature.
The restricted boltzmann machine sits at the boundary between classical approaches to feature extraction and more modern ML techniques while it also performs the best out of the listed classical algorithms.

Motivated by the performance of the boltzmann machine, we also studied an array of autoencoder based dimensionality reduction algorithms: a simple AE, a variational AE \citep{Kingma:2013hel}, a classifier-AE hybrid, and a end-to-end Sinkhorn AE \citep{deja2020end}.
Detailed descriptions of these methods are in App.\,\ref{sec:aes}.
Each of these algorithms imposes a different structure on the dimensionally reduced space.
We noticed that while the classifier AE performs extremely well, it is not robust.
Conversely, the variational AE is especially robust, but this leads to confounding between the classes in the generated data representation.
To achieve a middle ground between these two cases, we introduce the Sinkclass AE, a new architecture based on the Sinkhorn AE.

\subsection{The Sinkclass Autoencoder}
\label{sec:sinkclass}

\begin{figure*}
    \centering
    \includegraphics[width=0.9\textwidth]{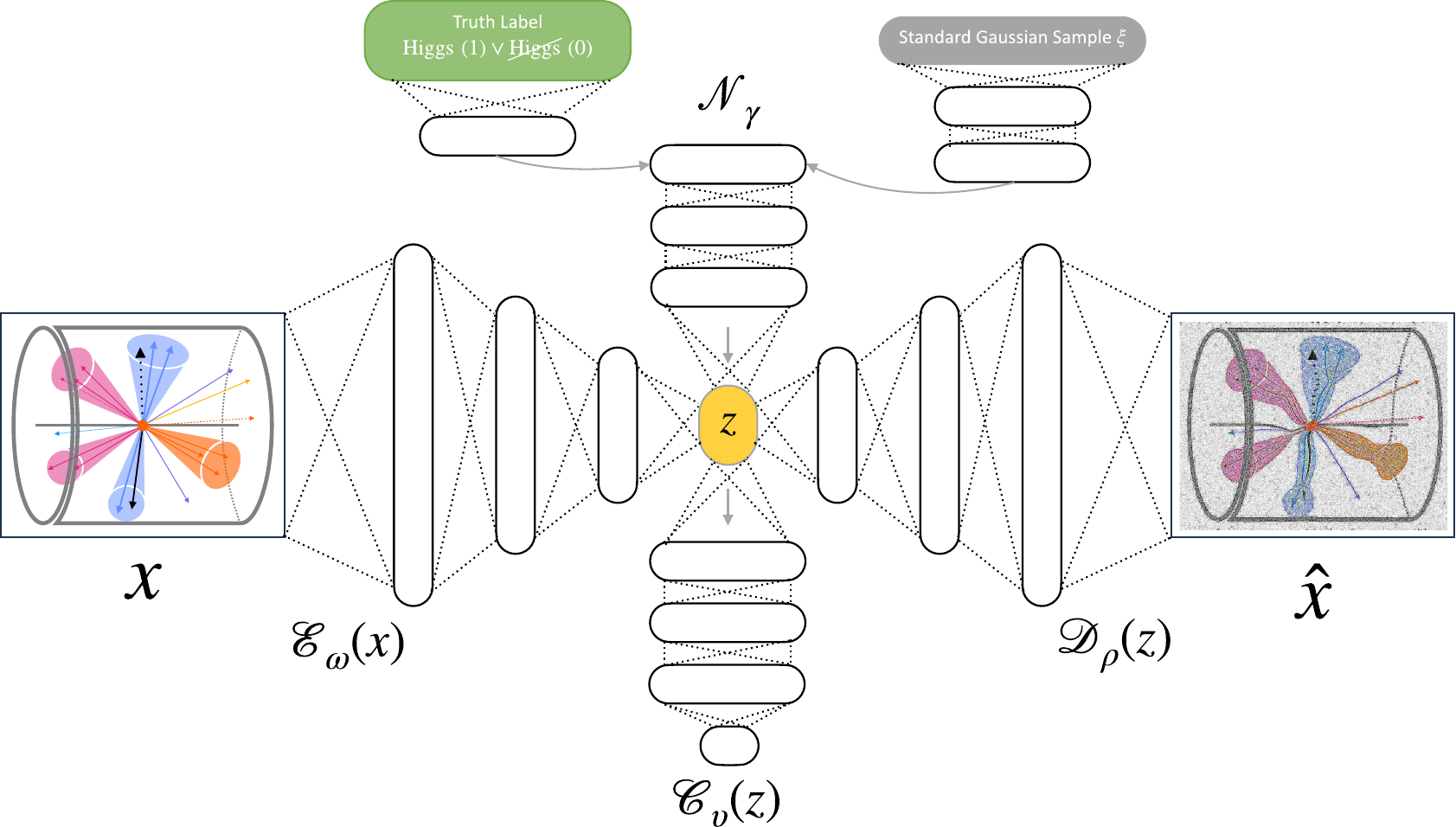}
    \caption{
    Schematic of the end-to-end Sinkhorn autoencoder with a classifier attached to its latent space; called the sinkclass autoencoder here. 
    The standard encoder-decoder structure is present.
    Furthermore, the truth labels, whether a sample contains a higgs boson or not, are fed to a one layer neural network.
    Simultaneously, a sample from a standard Gaussian distribution is given to a two layer neural network.
    The output of both networks are concatenated and given to a three layer network $\mathscr{N}_\gamma$ that connects to the latent space $z$ of the encoder.
    Similar to the variational autoencoder depicted in Fig.~\ref{fig:variationalAE}, the \ac{AE} component is trying to match the output of $\mathscr{N}_\gamma$ when computing $z$.
    The obtained $z$ is given to a conventional neural network classifier; its output is compared with the truth label and this information is used to compute an easier to classify $z$ for the next sample.
    Furthermore, the decoder $\mathscr{D}_\rho$ reconstructs the original data from $z$. 
    The goal of this network is to minimise the difference between $x$ and $\hat x$, to produce a $z$ that is close to the output of $\mathscr{N}_\gamma$, and that $z$ is easy to classify by $\mathscr{C}_\upsilon$
    }
    \label{fig:sinkclassAE}
\end{figure*}

Although the classifier \ac{AE} (App.\,\ref{sec:classifierAE}) increases the discriminative power of the latent space variables, the reconstruction performance is an order of magnitude worse than the one obtained by the variational~AE (App.\,\ref{sec:variationalAE}). 
\mbox{Moreover}, when the same classifier branch from the previous section is attached to the latent space of the variational AE and the resulting model is trained, the outcome does not increase either the classification or the reconstruction performance. 
The latent space feature distributions of the variational AE are constrained to each match a standard normal distribution by minimising the KL divergence between the two distributions; no distinction between signal and background is actually made. 
Instead, this model is rewarded for overlapping the two classes as closely as possible since the ensuing overall distribution would be closer to a standard Gaussian which decreases the $D_\mathrm{KL}$.
Conversely, the ideal latent space distribution for a classifier is represented by all signal events being at one end of the available range while all background events are at the other end. 
A latent space with such a shape would maximise $D_\mathrm{KL}$. 
Thus, the variational AE regularisation competes with class discriminability. 
One could use a (conditional) variational AE~\citep{kingma2014semi, NIPS2015_8d55a249} which is optimised towards approximating a latent bimodal distribution, each mode corresponding to a class, rather than a unimodal one.

However, such methods are quite unstable during training and they are hard to achieve convergence with, especially given data as complex as we are using here.
To solve all of these issues, a generalisation of the conditional variational AE, which we call the Sinkclass AE, is introduced. 
This new model, motivated by the latent space properties of the Sinkhorn \ac{AE}, represents a promising avenue for improving the discriminative power of the dimensionally reduced data representation while maintaining a good reconstruction. 
The Sinkclass AE architecture is shown in Fig.\,\ref{fig:sinkclassAE}.
In essence, the classifier neural network used in App.~\ref{sec:classifierAE} is attached to the latent space of the Sinkhorn~\ac{AE} described in App.\,\ref{sec:sinkhornAE}. 
Thus, instead of restricting the encoder to produce a fixed distribution, this distribution is approximated by using a conditional noise generator implemented with an additional deterministic neural network. 
Samples from a standard normal distribution are passed through the noise generator where they are encoded to follow the distribution of the $67$ feature input data in the latent space. 
Compared with the variational AE, the regularisation of the latent space is not as strict: the only requirement is that the Sinkhorn \ac{AE} latent space can be approximated with a non-linear mapping from a known distribution. 
Furthermore, an ordered target list filled with ones for every signal sample and zeroes for every background sample is also given to the conditional noise generator model. 
This encourages the encoding of the two data classes in different areas of the latent space and allows for a more natural, disentangled latent representation.
In addition, the separation between the classes is enforced even more through attaching an ML classifier to the latent space of the AE. 
The loss of the Sinkclass \ac{AE} is defined as
\begin{equation}
    \label{eq:sinkclassloss}
    \mathcal{L}_\mathrm{SCAE} = \alpha \mathcal{L}_\mathrm{SH} + \beta \mathcal{L}_\mathrm{BCE} + \mathcal{L}_\mathrm{MSE}
\end{equation}
where $\alpha$ and $\beta$ are weights in the range $[0, 1]$, $\mathcal{L}_\mathrm{SH}$ is the Sinkhorn loss, $\mathcal{L}_\mathrm{BCE}$ is the binary cross-entropy of the classifier, and $\mathcal{L}_\mathrm{MSE}$ is the mean squared error reconstruction loss. 
All these losses are detailed in App.\,\ref{sec:aes}.

The model is applied and optimised on the $67$-feature data set introduced in Sec.~\ref{sec:data}, after normalising the range of each feature distribution to $[0, 1]$.
As established in Eq.\,\ref{eq:sinkclassloss}, the Sinkclass AE has three loss components, which implies a complex loss landscape.
Therefore, the optimisation of its hyperparameters is done carefully in steps.
First, the learning rate and batch size of the Sinkclass \ac{AE} are tuned while fixing both the $\alpha$ and $\beta$ weights to~$1$. 
The~obtained values are $0.001$ for learning rate and $128$ for batch size. 
The structure of the encoder, noise generator, and decoder networks are the same as for the simple Sinkhorn AE presented in App.\,\ref{sec:sinkhornAE}.
Further, the weights $\alpha$ and $\beta$ are optimised once to obtain a model with the lowest unweighted $L_\mathrm{BCE}$ and once to obtain a model with the lowest unweighted $L_\mathrm{MSE}$. 
The former search returned the values $\alpha = 0.2$ and $\beta = 0.02$, while the latter returned $\alpha = 0.0008$ and $\beta = 0.9$. 
Despite the complex loss function, the model trains and converges.
Additionally, it generates a lower dimensional representation of the input data that is suitable to be used as input for a currently deployable QML classifier.
The quantum support vector machine is chosen as our QML classifier here, given its simplicity.

\section{Quantum Support Vector Machine}
\label{sec:qsvm}

A \ac{SVM} is a supervised \ac{ML} architecture and also a kernel method~\citep{svmVapnik}. 
It~operates by constructing a hyperplane as a linear decision boundary between two classes of data points. 
By~transforming data points into a higher-dimensional feature space, \ac{SVM}s can effectively tackle intricate classification tasks requiring non-linear decision boundaries between the classes. 
The efficacy of \ac{SVM}s stems from the utilisation of the \textit{kernel trick}, which uses a similarity measure between data points, known as the kernel, to circumvent the need for explicit computation of the transformation to the higher-dimensional feature space. 

\begin{figure*}[th]
    \centering
    \includegraphics[width=\textwidth]{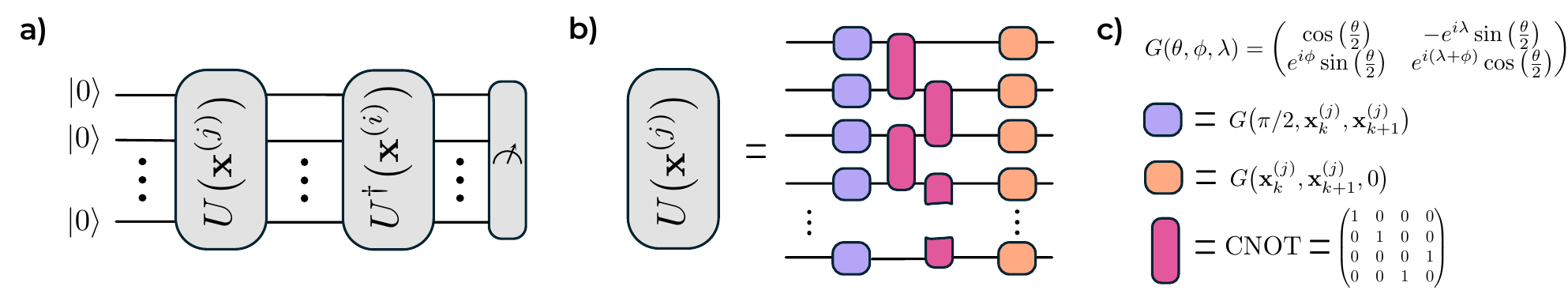}
    \caption{\textbf{a)} The quantum kernel circuit, given a data encoding circuit $U$, from which the quantum kernel values are extracted via the measurement of the expected value defined in Eq.~\ref{eq:quantum_kernel}. \textbf{b)} The architecture of the data encoding circuit from~\citet{Belis:2023atb}. \textbf{c)} The components of the data encoding circuit, where $G(\theta,\phi,\lambda)$ is the universal one-qubit gate, $\mathbf{x}^{(j)}_k$ is the $k$-th element of the input data vector $\mathbf{x}^{(j)}$}
    \label{fig:qsvm}
\end{figure*}


The dual formulation of the \ac{SVM} optimisation problem is given by:
\begin{align}
&\min_{\boldsymbol{\alpha}}\left(\sum_{i=1}^N \alpha_i - \frac{1}{2}\sum_{i,j=1}^N \alpha_i \alpha_j y_i y_j\,k(\mathbf{x}^{(i)}, \mathbf{x}^{(j)})\right), \label{eq:svm_dual_final}\\
&\text{subject to:}\; \sum_{i=1}^N \alpha_i y_i = 0,\quad 0 \leq \alpha_i \leq C,\; \forall i, \label{eq:svm_dual_C}
\end{align}
where $k(\mathbf{x}^{(i)}, \mathbf{x}^{(j)})$ is the kernel function, $N$ is the number of training samples, $\alpha_i$ are the optimisation variables that define the decision boundary, $y_i, y_j \in \{-1, 1\}$ are the labels corresponding to the input vectors $\mathbf{x}^{(i)}$ and $\mathbf{x}^{(j)}$, respectively, and $C$ is a regularisation hyperparameter. The values of $\alpha_i$ are obtained by minimising Eq.~\ref{eq:svm_dual_final}.

For the \ac{QSVM}, kernel values are assessed within a Hilbert space of states~\citep{Havlicek2019,Schuld_2019}. 
\ac{QSVM}s equipped with well-chosen quantum kernels can outperform their classical counterparts in some applications \citep{Liu2021rigorous, muser2023, Belis:2023atb, molteni2024}. 
Initially, classical features $\mathbf{x} \in \mathcal{X}$, which in our case are the dimensionally-reduced samples, are encoded into the quantum state space $\mathcal{F}$ via a feature map $\varphi: \mathcal{X} \rightarrow \mathcal{F}$, defined as:
\begin{equation}
\mathbf{x} \rightarrow \varphi(\mathbf{x})\equiv\rho(\mathbf{x}) = \ket{\phi(\mathbf{x})}\bra{\phi(\mathbf{x})},
\end{equation}
where $\rho$ is the density matrix.
The quantum feature map, also referred to as the data encoding circuit, is implemented via a parameterised unitary transformation $U(\mathbf{x})$ applied to a fixed reference state, such as $\ket{\phi(\mathbf{x})} = U(\mathbf{x}) \ket{0}$. 
Subsequently, the quantum kernel between two data points $\mathbf{x}^{(i)}$ and $\mathbf{x}^{(j)}$ is computed as the overlap between two states that encode these data points via the Hilbert-Schmidt inner product
\begin{equation}
k(\mathbf{x}^{(i)}, \mathbf{x}^{(j)}) = \mathrm{Tr}\left[ \rho(\mathbf{x}^{(i)})\cdot \rho(\mathbf{x}^{(j)}) \right] = \left| \braket{\phi(\mathbf{x}^{(i)})|\phi(\mathbf{x}^{(j)})} \right|^2.    
\label{eq:quantum_kernel}
\end{equation}
The overlap between the encoded quantum states corresponds to the probability of measuring the all-zero state in the computational basis. To estimate this probability on a quantum computer we measure the final state of all qubits $R$ times counting the frequency of the $0^{n_q}$ bit-string, where $n_q$ is the number of qubits.

Various strategies exist for encoding classical data into a quantum state~\citep{schuld_qml_is_kernel2021}. 
We adopt the kernel introduced in~\citet{Belis:2023atb} and depicted in Fig.~\ref{fig:qsvm}. 
Optimising the quantum kernel circuit to achieve a notable improvement in classification performance for the specific \tth classification task is outside the scope of this study; see ~\citet{Altares-López_2021, altares_lopez2024, incudini2023automatic} for that. 
The goal of our work is to investigate the dimensionality reduction methods presented in Sec.~\ref{sec:dimred} and the effect they have on the classification performance of the same well-understood quantum classifier, i.e., the QSVM with a fixed feature map. 

The data encoding circuit we use encodes two input features per qubit via unitary rotation gates and implements nearest-neighbours entanglement between qubits via CNOT gates (Fig.~\ref{fig:qsvm}b). 
The unitary rotation gates are repeated, but with the rotation angles permuted such that they correspond to different input features from the first layer of rotations (see Figure \ref{fig:qsvm}c). 
This repetition of gates and permutation of encoding angles increases the expressibility of the quantum circuit \citep{Belis:2023atb}.

We simulate the \ac{QSVM} on a classical processor using Qiskit~\citep{qiskit} to investigate the impact of the various reduced representations on its performance in an ideal environment, free from quantum hardware noise. 
For each data set produced by the studied dimensionality reduction methods, we perform a grid search over the regularisation hyperparameter $C\in\{10^{-3}, 10^{-2}, 10^{-1}, 10^0, 10^1\}$ of the \ac{QSVM}, as defined in Eq.~\ref{eq:svm_dual_C}, converging to $C=1$.

\section{Results and Discussion}\label{sec:results}

All feature extraction methods employed in this study reduce the size of the data from $D=67$ to $D^*=16$.
The~data is processed as described in Sec.~\ref{sec:data} and minmax normalisation is applied to map the range of all the 67 features to $[0, 1]$.
After this preprocessing stage, the trainable dimensionality reduction algorithms, are trained on $1.15\times 10^6$ samples.

A validation data set of $1.44\times 10^5$ samples is used to optimise and validate the hyperparameters of the model.
Furthermore, a test data set of $1.44\times 10^5$ samples is used to report the final losses and AUCs of the models.
The \ac{QSVM} is trained on a subset of 600 events from the dimensionally reduced test data set.
Finally, the performance of the \ac{QSVM} is determined by running it on another subset of the dimensionally reduced test data set containing 3600 events.
The uncertainty pertaining to the size of the test data is approximated by splitting the test data set into five equal subsets, running our algorithms independently on each subset, and then averaging the results.
All the data sets mentioned above have equal signal and background samples.
The results obtained by following this paradigm are shown in Tab.\,\ref{tab:classicalTable} and Tab.\,\ref{tab:AEtable}.

\begingroup
\setlength{\tabcolsep}{16pt} 
\renewcommand{\arraystretch}{0.5} 
\begin{table*}[htb!]

\centering
\resizebox{\textwidth}{!}{%
\begin{tabular}{l c c c}
\toprule\toprule
Method                                  & Feature Extraction Type & QSVM AUC      \\ \midrule\midrule
\ac{ICA}                               & Linear                  & 0.528 ± 0.006 \\ \midrule
\ac{NMF}                               & Linear                  & 0.599 ± 0.013 \\ \midrule
\ac{PCA}                               & Linear                  & 0.541 ± 0.015 \\ \midrule
\ac{LLE}                               & Non-linear: Manifold Learning       & 0.533 ± 0.014 \\ \midrule
\ac{SE}                                & Non-linear: Manifold Learning       & 0.526 ± 0.013 \\ \midrule
\ac{RBM}                               & Non-linear: Neural Network          & 0.651 ± 0.016 \\ \bottomrule
\end{tabular}%
}
\caption{
The classification performance of the \ac{QSVM} on dimensionally reduced representations of the data obtained by using some of the most customary classical methods.
Each of the feature extraction algorithms is described in App.~\ref{sec:classicalfe}; their type is also specified, from simple linear methods to neural networks.
All the models shown in this table take input with $D=67$ features and compute a reduced representation of $D^*=16$ features.
The uncertainties on the \ac{AUC} are computed by running the \ac{QSVM} on five reduced test set subsets and averaging the results.
The \ac{RBM}, which is the closest classical method to an \ac{AE} shows the best results.
}
\label{tab:classicalTable}
\end{table*}
\endgroup

\begingroup
\setlength{\tabcolsep}{10pt} 
\renewcommand{\arraystretch}{0.5} 
\begin{table*}[htb!]
\centering
\resizebox{\textwidth}{!}{%
\begin{tabular}{l c c c c c}
\toprule\toprule
Autoencoder                 & Optimisation    & MSE Loss $\times 10^{-4}$ & BCE Loss & Classifier AUC    & QSVM AUC        \\ \midrule\midrule
Vanilla                     & -               & 4.77                      & -        & -                 & 0.56$\pm\,0.01$ \\ \midrule
Variational                 & MSE             & 4.49                      & -        & -                 & 0.56$\pm\,0.02$ \\ \midrule
\multirow{2}{*}{Classifier} & MSE             & 5.47                      & 0.63     & 0.700$\pm$\,0.001 & 0.56$\pm\,0.02$ \\ \cmidrule(l){2-6} 
                            & BCE             & 62.97                     & 0.61     & 0.734$\pm$\,0.002 & 0.72$\pm\,0.01$ \\ \midrule
Sinkhorn                    & MSE             & 9.65                      & -        & -                 & 0.51$\pm\,0.01$ \\ \midrule
\multirow{2}{*}{Sinkclass}  & MSE             & 26.41                     & 0.65     & 0.642$\pm$\,0.003 & 0.50$\pm\,0.01$ \\ \cmidrule(l){2-6} 
                            & BCE             & 24.69                     & 0.61     & 0.734$\pm$\,0.002 & 0.74$\pm\,0.01$ \\ \bottomrule
\end{tabular}%
}
\caption{
The classification performance of the \ac{QSVM} on the reduced representations of the data computed with \ac{AE} methods.
The models presented in this table are described at length in App.~\ref{sec:aes}.
The optimisation column refers to how the hyperparameters of that \ac{AE} were optimised, whether to minimise the \ac{MSE} or the \ac{BCE} loss of the validation data .
The losses are reported on the test data, along with the classifier \ac{AUC} wherever applicable.
The \ac{QSVM} \ac{AUC} is computed on a subset of this test data, since simulation of quantum circuits is computationally expensive.
The uncertainties on the \ac{AUC} are computed by running the \ac{QSVM} on five test data subsets and averaging the results.
The Sinkclass \ac{AE}, optimised for \ac{BCE} minimisation leads to a \ac{QSVM} pefrormance that is competitive with state of the art methods for this particular data set~\citep{reissel21}.
}
\label{tab:AEtable}
\end{table*}
\endgroup

The results of the classical dimensionality reduction methods are shown in Tab.~\ref{tab:classicalTable}.
The \ac{AUC} obtained by the \ac{QSVM} is shown along with the type of feature extraction algorithm for each of the six employed models described in Sec.~\ref{sec:conventional} and described in App.\,\ref{sec:classicalfe}.
Out of the linear methods, \ac{NMF} performs the best, showing a significant 13\% improvement over the \ac{PCA} algorithm.
The manifold learning models that we investigated, namely LLE and SE, did not yield good results for our data set, but might show better performance in problems where local correlations are more important.
Among the non-linear feature extraction methods, the \ac{RBM} algorithm shows the best performance.
The RBM model is relatively similar to modern deep-learning \ac{AE}s; thus, observing an increase in performance is consistent with our expectations and motivates the study of AEs.
Consequently, multiple \ac{AE} models are studied and the results are show in Tab.~\ref{tab:AEtable}.
Notice, the performance of \ac{AE} models is lower than that of the RBM, i.e., increased model complexity does not guarantee better results.

The different losses and classifier AUCs, for the models that incorporate a classifier $\mathscr{C}_\upsilon$ in their architecture, are displayed in Tab.\,\ref{tab:AEtable}.
The models where the \ac{AE} includes a classification task besides its reconstruction task exhibit the most discriminative dimensionally reduced spaces.
Specifically, the sinkclass \ac{AE} optimised for best classification performance displays the best discriminative power in its latent space out of all our architectures. 

Additionally, the Sinkclass possesses a relatively good reconstruction ability and a regularised latent space.
The~former is reflected in the relatively low classifier binary cross-entropy (BCE) loss while the latter in the reconstruction mean square error (MSE) loss.
The AUCs of the QSVM classifiers from the \ac{AE}s that include a classical classifier are better than the baseline: a neural network classifier applied on all features \citep{Belis:2021zqi}.
The \ac{QSVM} applied to the latent space of the Sinkclass \ac{AE} produces a comparable performance to that of classical state-of-the-art neural networks~\citep{reissel21}; albeit the increase in performance with respect to the \ac{AE}'s own classifier is marginal.
Both the Classifier \ac{AE} and Sinkclass \ac{AE} exhibit an \ac{MSE} loss on the order of $10^{-4}$, with Sinkclass \ac{AE} having less than half the \ac{MSE}. 
When trained to optimize \ac{BCE} loss (signal-background separation in the latent space), both models learn similarly representative latent distributions of the input data. 
However, the \ac{QSVM} performs slightly better on the compressed data from the Sinkclass AE compared with the performance reported by the classifier branches of the investigated AEs, indicating that its latent space aligns better with the quantum kernel.

\section{Conclusion}
\label{sec:conclusion}

The application of \ac{QML} algorithms to complex problems is currently impeded by the high number of features required to describe such a problem.
There exist two obvious ways of proceeding: either wait for current \ac{QML} hardware to increase in capacity and availability or use dimensionality reduction techniques.
Choosing to apply dimensionality reduction techniques immediately poses the question of which dimensionality reduction method one should use for their problem.
Currently, no general answer exists: each data set is different, and furthermore, multiple problems can be defined on a single data set.
However, we perform a systematic study pertaining to a diverse set of dimensionality reduction problems, applied to a realistic data set, and extract a few heuristics.
These heuristics can then be applied to other high-dimensional data sets for enabling QML use.

The lower-dimensional representations of our chosen \tth data set, generated by the aforementioned algorithms, are given to a \ac{QSVM} for binary classification.
The performance of the \ac{QSVM}, represented by its \ac{AUC}, is shown in Tab.~\ref{tab:classicalTable} and Tab.~\ref{tab:AEtable}.
We observe that the \ac{AE} based methods present, in general, better latent spaces for \ac{QSVM} classification, although more complex and less interpretable than their classical method counterparts.
Furthermore, for the \ac{AE}s, we show that learning to perform the dimensionality reduction task at the same time as the classification task leads to a dimensionally reduced set of features that is more discriminative.
In essence, the lower-dimensional representation produced by these models presents a better separability between the classes.
The architecture that exhibits all of these attributes for our data is the introduced Sinkclass AE.
However, we propose two heuristics for dimensionality reduction in the context of \ac{QML} applications: \ac{AE} based dimensionality reduction is likely to give better results than conventional methods and one should combine dimensionality reduction with the downstream task, e.g., classification.

As we show in this work, the latter does not need to be performed by using the exact architecture of the downstream task,  but any other architecture that addresses the task successfully. 
Thus, we avoid situations in which this process is intractable due to how expensive the training of QML architectures is.

\section*{Data availability}\label{sec:data_availability}
\noindent
The data sets used for this study are publicly available on Zenodo~\citep{tthbb_data}.

\section*{Code availability}\label{sec:code}
\noindent
The code developed for this paper is available publicly in the GitHub repository: \url{https://github.com/bb511/dimred4qsvm}.

\section*{Acknowledgements}
The authors would like to thank Elías F. Combarro for the helpful discussions.

\section*{Funding}
V.B. is supported by an ETH Research Grant (grant no.~ETH C-04 21-2). P.O. is supported by Swiss National Science Foundation Grant No.~PZ00P2\_201594.  M.G. and S.V. are supported by CERN through the Quantum Technology Initiative. F.R. acknowledges financial support by the Swiss National Science Foundation (Ambizione grant no. PZ00P2$\_$186040).

\newpage
\clearpage
\bibliography{main}

\onecolumngrid
\newpage
\section*{Appendices}
\vspace{5mm}
\appendix

Each of the models detailed in App.\,\ref{sec:classicalfe} and App.\,\ref{sec:aes} is implemented and applied to the $67$-feature data set introduced in Sec.~\ref{sec:data}, after normalising the range of each feature distribution to $[0, 1]$.
We define a data set from the input feature space as matrix $\textbf{X}\in\mathbb{R}^{D\times M}$ with $M$ samples $\textbf{x}^{(i)}\in\mathbb{R}^D$ of $D=67$ features each. 
We aim to find a feature extraction method $f:\mathbb{R}^D\to\mathbb{R}^{D^{*}}$ where $1\leq D^{*}< D$ such that the information retained in $\widehat{\textbf{X}}\in\mathbb{R}^{D^*\times M}$ is sufficiently expressive to allow binary classification using the resulting feature~space. 
In~this~study, we fix $D^{*}=16$.

\vspace{5mm}
\section{Classical Feature Extraction}
\label{sec:classicalfe}

\subsection{Principal Component Analysis}
\label{sec:PCA}

\ac{PCA} is one of the most frequently used methods for feature extraction as it constitutes the lowest-error linear dimensionality reduction method available \citep{osti_15002155}. 
Thus, it is often the default choice when a more nuanced feature extraction method selection is unnecessary for the downstream application.
Conceptually, PCA defines the new feature dimensions as a set of new axes that span the original feature space, where the best $D^*$ dimensions are kept.  
These axes, called principal components, are orthogonal to each other and are constructed incrementally such that the projection of the data points onto the next axis maximises the remaining variance.
Thus, the variance of the original data preserved in the reduced data set is maximal for the given number of dimensions. 
Alternatively, one can set a target variance share of the original data to achieve with PCA and choose the number of principal components. 
The projected values are shifted such that they are centered around zero \citep{osti_15002155}.

The principal components are mathematically defined as follows \citep{bishop2006pattern}.
Consider $\textbf{S}$ to be the covariance matrix $\textbf{S} = \frac{1}{M} \sum^M_{i=1} (\textbf{x}^{(i)} - \bar{\mathbf{x}})(\textbf{x}^{(i)} - \bar{\mathbf{x}})^\top$. The magnitude of the variance projected onto a dimension $\textbf{u}_1$ is given by $\textbf{u}^\top_1 \textbf{S} \textbf{u}_1$, where $\textbf{u}_1$ is the first principal component, a $D$-dimensional vector. 
The maximum of the variance that may be obtained while constraining $\textbf{u}_1$ to be of unit length is an eigenvalue $\lambda_i$.
Thus, the ﬁrst principal component is the eigenvector $\mathbf{v}_i$ that corresponds to the largest eigenvalue, 
$\textbf{u}_1 = \textbf{v}_i$ s.t. $i = \arg \max_{j \in [D]} \lambda_j$.
For the next principal component $\textbf{u}_2$, the covariance matrix $\textbf{S}'$ of the data minus its projection onto the first principal component $\textbf{X}-\textbf{u}_1\textbf{u}^\top_1\textbf{X}$ is considered. The vector $\textbf{u}_2$ is chosen to maximise the magnitude of $\textbf{S}'$ projected onto it, $\textbf{u}^\top_2\textbf{S}'\textbf{u}_2$, the solution of which is the eigenvector $\textbf{v}_k$ of $\textbf{S}$ corresponding to the second largest eigenvalue $\lambda_k$.
The reduced feature space is consequently spanned by the $D^*$ largest-eigenvalue eigenvectors in decreasing order $\textbf{u}_1, \dots, \textbf{u}_{D^*}$. 
The eigenvectors can be found using an eigensolver, where we use randomised singular value decomposition \citep{halko2011finding}. 
The~share of variance contained in each principal component can be computed by dividing its corresponding eigenvalue by the sum of all eigenvalues. 
The dimensionally reduced data set is then $\widehat{\textbf{X}} = \textbf{U}^\top \textbf{X}$ where $\textbf{U}=[\textbf{u}_1, \dots, \textbf{u}_{D^*}]$.  



\subsection{Independent Component Analysis}
\label{sec:ICA}

\ac{ICA} assumes that the observed data is a linear combination of statistically independent latent components which can be reconstructed~\citep{6918213}. 
Conceptually, ICA is similar to PCA with the difference that the axes that represent the new features are not restricted to be orthogonal; they are chosen such that they encode variance independently of the other components. 
To identify these components either the mutual information between the components is minimised or their distinctiveness from a Gaussian distribution is maximised \citep{6918213}.

We assume $D^*$ independent source signals (components) mixed in the data that is observed in a total of $D$ ways. The mixing is determined by a matrix $\textbf{A}\in\mathbb{R}^{D\times D^*}$ such that $\textbf{X}=\textbf{A}\textbf{S}$ where $\textbf{S}\in\mathbb{R}^{D^*\times M}$ are the source signals. 
The~objective is to find the unmixing matrix to $\textbf{A}$, $\textbf{W}\in\mathbb{R}^{D^*\times D}$, so that the reconstructed source data is $\widehat{\textbf{X}}=\textbf{W}\textbf{X}$. 
This~is achieved by optimising one of multiple metrics of independence in the reconstructed source signals $\widehat{\textbf{X}}$ with respect to $\textbf{W}$, e.g., maximising the negentropy for non-Gaussianity of the source distributions or minimising mutual information for shared source information. 
However, since both $\textbf{A}$ and $\textbf{S}$ are unknown, specific ambiguities such as the order of components may limit how well the reconstruction is performed~\citep{hyvarinen1999independent}. 
We use the iterative FastICA implementation \citep{hyvarinen2000independent} with negentropy as target approximated via a cubic $G$ function.

\newpage
\subsection{Locally Linear Embedding}
\label{sec:LLE}

\ac{LLE} is a non-linear transformation method that aims to preserve the non-linear structure of data points in the lower-dimensional space. 
This is different from the linear methods shown hitherto.
Non-linear structures are captured by local linear ﬁts in the higher-dimensional space that should persist locally in the lower-dimensional space \citep{roweis2000nonlinear}. 

Formally, each data point $\textbf{x}^{(i)}$ in the $D$-dimensional space is assigned a set of $K$ neighbouring data points and is represented via a linear combination of these points.
A~point $\textbf{x}^{(i)}$ with a set of neighbour indices $\mathcal{J}_i\subset\mathbb{N}$, $|\mathcal{J}_i|=K$, can be written as: $\textbf{x}^{(i)} \approx \sum_{j\in \mathcal{J}_i} w_{i,j}\textbf{x}^{(j)}$. 
The~weights obey $\sum_{j\in\mathcal{J}_i}w_{i,j}=1$ for a data point, and $w_{i,j\notin \mathcal{J}_i} = 0$ for non-neighbouring pairs of points.  
\mbox{Using} the weights from the higher-dimensional space in the lower-dimensional space gives rise to lower-dimensional representations with the same local structure. 

The algorithm is ﬁt in three steps. First, indentify the $K$ nearest neighbours of each data point through an algorithm such as $K$-nearest neighbours or ball trees.
Second, the weights matrix $\textbf{W}\in\mathbb{R}^{M\times M}$ is determined through minimising
\begin{equation}
\label{eq:6}
\tilde{\textbf{W}} = \arg \min_\textbf{W} \sum^M_{i=1}\bigg\| \textbf{x}^{(i)} - \bigg( \sum_{j\in \mathcal{J}_i} w_{i,j}\textbf{x}^{(j)}\bigg ) \bigg\|^2  
\end{equation}
which represents a reconstruction error.
This is analytically solvable as a least-squares \mbox{problem}.
Third, the corresponding $D^*$-dimensional representations $\hat{\textbf{x}}^{(i)}$ are computed by minimising Eq.\,\ref{eq:6} with respect to $\widehat{\textbf{X}}$ while replacing $\textbf{x}^{(i/j)}$ with $\hat{\textbf{x}}^{(i/j)}$ and ﬁxing $\textbf{W} = \tilde{\textbf{W}}$. 
For this optimisation, an eigenvalue problem is solved \citep{roweis2000nonlinear}.

Among the advantages of \ac{LLE} is the limited number of hyperparameters, such as the number of neighbours $K$. More neighbours per data point usually lead to a smoother embedding, whereas fewer neighbours yield a higher sensitivity to small local structures. 
We set $K=12$.
The algorithm that determines the nearest neighbours, as well as the employed eigensolver, constitute two additional choices. 
Here,~a~dense eigen-decomposition 
is used, along with a 
brute force $K$-nearest neighbours search.
To avoid the issue of rank deﬁciency if the number of neighbours per data point is higher than the number of original features, $\textbf{W}$ is regularised.




\subsection{Spectral Embedding}
\label{sec:SE}

SE \citep{ng2001spectral, belkin2003laplacian, von2007tutorial} is another non-linear method that assumes the data points in the higher-dimensional space to be sampled from a lower-dimensional manifold. 
The aim is to reveal properties about this manifold in order to represent the data points in its lower-dimensional space. 
SE uses a similarity graph that captures the pairwise similarities of data points in the original space; the Laplacian of this graph has eigenvectors that are used for the embedding.  

First, the similarities are computed according to one of multiple similarity metrics $s(\textbf{x}^{(i)}, \textbf{x}^{(j)}) = s_{i,j}\geq 0$, where more similar data points give rise to larger values. 
\mbox{Second}, an undirected similarity graph $G$ is constructed which consists of all $M$ data points as vertices $\{v_i\}_{i=1}^M$ and edges $e_{i,j}$ connecting two vertices $v_i, v_j$. 
There exist a lot of different strategies to construct the graph, of which we consider three.
One option is to connect all vertices and have the similarity measure $s$ determine the extent of the relationship between data points by weighting their edges. 
Another option is connecting vertex pairs with a similarity value larger than some threshold with a uniform weight. 
Alternatively, only vertex pairs where one data point is in the $K$ nearest neighbours of the other may be connected. 
The similarity graph gives rise to an $M\times M$ symmetric weighted adjacency matrix $\textbf{W} = (w_{i,j})_{i,j=1,\ldots,M}$, where $w_{i,j}$, consistent with the chosen strategy, may assume $s_{i,j}$ or the result of some other function for connected vertices and zero for non-connected vertices. 
Additionally, the similarity graph is also described by a degree matrix deﬁned as $\textbf{D} = \mathrm{diag}(d_1, \dots, d_M )$ where $d_i = \sum^M_{j=1} w_{i,j}$, the sum over the weights of the edges for a vertex $v_i$.
Then,  $\mathbf{L}\in\mathbb{R}^{M\times M}$, the so-called graph Laplacian matrix, is determined according to one of \mbox{two~methods}
\citep{belkin2003laplacian}. 
The~method used in this work consists of defining $\textbf{L}$ as a normalised and positive semi-definite matrix, i.e., $\textbf{L} = \textbf{I} - \textbf{D}^{-1}\textbf{W}$, where $\textbf{I}$ denotes the identity matrix~\citep{ng2001spectral}.
If the graph $G$ is not connected then $\textbf{L}$ can be constructed as a block diagonal matrix from the Laplacians of each connected block of the graph \citep{von2007tutorial}.
The Laplacian $\textbf{L}$ of a similarity graph $G$ captures geometric information about the locality of the data points. 

Lastly, consider the $D^*$ eigenvectors $\textbf{u}_1 , \dots , \textbf{u}_{D^*}$ of $\textbf{L}$, corresponding to the $D^*$ smallest eigenvalues $\lambda_1 , \dots , \lambda_{D^*}$ while excluding $\lambda_0 = 0$. 
These provide an embedding of the data points in the lower-dimensional space $\mathbb{R}^{D^*}$. 
After establishing the graph and the Laplacian, the ﬁtting is concluded by constructing the eigenvector matrix $\textbf{U} = [\textbf{u}_1 , \ldots , \textbf{u}_{D^*}] \in \mathbb{R}^{M\times D^*}$.

Thus, the eigenvectors $\textbf{U}$ of $\textbf{L}$ constitute the embedding of the data set $\textbf{X}$. 
Specifically, the transformation of an original data point $\textbf{x}^{(i)}$ is given by the $i$-th components of the eigenvectors $\hat{\textbf{x}}^{(i)} = (\textbf{U}_{i,*})^\top = [u_{1i}, \dots, u_{D^*i}]^\top$. 
The conﬁgurable hyperparameters for this dimensionality reduction algorithm are the eigensolver algorithm and the method used to construct the similarity graph. \
Here, we construct the graph as a symmetric nearest neighbour graph with binary weights and $K=9$ neighbours.


\subsection{Non-Negative Matrix Factorisation}
\label{sec:NME}

In \ac{NMF}, the high-dimensional input data matrix $\textbf{X}$ is factorised into two lower-dimensional matrices of complementary sizes~$\textbf{W},\textbf{H}$.
The input and its factorisation matrices are constrained to consist only of non-negative values. 
The two matrices $\textbf{W},\textbf{H}$ are obtained via minimisation of a reconstruction error. 
The non-negativity constraint in this optimisation process leads to sparsity \citep{6165290}.


Formally, in \ac{NMF}, the non-negative data matrix $\textbf{X} \in \mathbb{R}^{D\times M}_{0^+}$ is factorised into a non-negative basis matrix called  $\textbf{W} \in \mathbb{R}^{D\times D^*}_{0^+}$ and a non-negative coeﬃcient matrix $\textbf{H}\in \mathbb{R}^{D^*\times M}_{0^+}$ such that $\textbf{X} \approx \textbf{W}\textbf{H}$. 
These matrices are fit to minimise the distance between the data matrix and the approximation according to a distance metric, such as the squared Frobenius norm: $\|\textbf{X} - \textbf{W}\textbf{H}\|^2_F$. 
The optimisation algorithm may be one of diﬀerent numerical solvers such as coordinate descent or the multiplicative update ~\citep{6165290}.
Diﬀerent methods exist to initialise the mentioned matrices.
Thus, the tunable hyperparameters are given by the distance metric, the regularisation conﬁguration, the optimisation algorithm, and the approach of the initial construction of the matrices. 
In this work we choose KL divergence, $L_1$ regularisation of $\textbf{W}$ and $\textbf{H}$, the multiplicative update method, and the NNDSVDA initialisation.

After fitting, the $D$-dimensional column vectors of the basis matrix, $\{\textbf{w}_{*,k} \}_{k=1}^{D^*}$, are the $D^*$ basis vectors for which the $D^*$-dimensional column vectors of the coefficient matrix $\{\textbf{h}_{*,i}\}_{i=1}^M$ are the respective coefficients of~the~$M$~data~points. 
Each $D$-dimensional original data point $\textbf{x}^{(i)}$ is reconstructed as a linear combination of the $D^*$ basis vectors such that $\textbf{x}^{(i)} = \textbf{W}\textbf{h}_{*,i}$. 
The $D^*$-dimensional coeﬃcients vector is the dimensionally reduced approximate representation of the corresponding $D$-dimensional data point. 
Hence, $\widehat{\textbf{X}}=\textbf{H}$ is the reduced representation \citep{paatero1994positive}. 

Beyond the basic \ac{NMF} employed here, further \ac{NMF} variants exist that can be grouped into constrained \ac{NMF}, structured \ac{NMF}, and generalised \ac{NMF}. 
A description of these methods is available in \citet{6165290}.


\subsection{Bernoulli Restricted Boltzmann Machine}
\label{sec:RBM}

The \ac{RBM} is an unsupervised shallow artiﬁcial neural network and generative model that aims to reconstruct the given input by maximising the likelihood of the training data~\citep{smolensky1986information,hinton2002training}. 
It consists of two fully connected layers of diﬀerent unit count where the ﬁrst layer has $D$ units and the second so-called hidden layer has $D^*$ units. 
It is restricted compared to classical Boltzmann Machines \citep{ackley1985learning} since each unit of one layer is connected to all units of the other layer but not to the nodes of the same layer. 


The input units $\{v_k\}_{k=1}^D$ are connected with the hidden units $\{h_l\}_{l=1}^{D^*}$ as defined by the symmetric weights $w_{k,l}$. 
Each unit $v_k/h_l$ has a bias $b_{k|l}^{(v|h)}$. The activation of a hidden unit, that is, its conditional probability of being activated given the activations of the input units is defined as a logistic sigmoid over the sum of all connection-weighted values of the incoming units and its own bias: $P(h_l=1|v_1,\ldots,v_D) = \sigma\left (b_l + \sum_{k=1}^D w_{k,l}v_k \right )$. This holds analogously for the probabilities of the input units being activated given the hidden activations. The specific type of \ac{RBM} used here is called Bernoulli \ac{RBM}s, since the input units $\textbf{x}^{(i)}=\textbf{v}=[v_1,\ldots,v_D]$ assume binary or $[0, 1]$-normalised real values.

The weights $\textbf{W}$ and biases $\textbf{b}^{(v)},\textbf{b}^{(h)}$ are fit in a gradient-based optimisation according to the persistent contrastive divergence algorithm \citep{tieleman2008training,hinton2002training}, using approximate gradients of the likelihood of the data with respect to the parameters. 
The following sequence forms an update step: the hidden units are initialised with evaluations of their activations $\textbf{h}$ using the input activations of a training sample $\textbf{v}$ and initial parameters; then, the so-called reconstructed input units' activations $\textbf{v}'$ are evaluated according to their distributions defined by $\textbf{h}$ and the reconstructed hidden layer activations $\textbf{h}'$ are evaluated based on $\textbf{v}'$. 
The update delta is then given by $\Delta \textbf{W}=\alpha \left( \textbf{v}\textbf{h}^\top - \textbf{v}'{\textbf{h}'}^\top\right )$ and $\Delta \textbf{b}^{(z \in \{v,h\})}=\alpha \left( \textbf{z} - \textbf{z}' \right)$. 
This loop terminates when a suﬃciently low reconstruction error is reached. In the persistent version \citep{tieleman2008training} of the same algorithm used in this study, the chain of reconstructions is continued from the output of the previous step obtained from the previous sample, rather than starting anew for each step. 
The reduced data set is then obtained by passing the input data to the visible layer and evaluating the activations of the hidden layer: 
$$
\hat{\textbf{x}}^{(i)}=[P(h_1=1|v_1,\ldots,v_D),\ldots,P(h_{D^*}=1|v_1,\ldots,v_D)]^\top,
$$ 
where $v_k=x^{(i)}_k$.

Despite appearing less frequently in the current dimensionality reduction literature, the Bernoulli \ac{RBM} is considered here because of its similarity to autoencoders~\citep{doi:10.1126/science.1127647}. 
The tunable hyperparameters in this method are the batch size, the learning rate, and the number of training epochs. 
Here we use $100$, $0.05$, and $10$, respectively.
These parameters enter the optimisation process described above.

\section{Autoencoder Methods}
\label{sec:aes}

The \ac{AE} is a well-known machine learning architecture~\citep{lecun1987phd, ballard1987modular, hinton1993autoencoders}, commonly used nowadays for dimensionality reduction in many fields of science.
It is similar to the \ac{RBM}, introduced in App.~\ref{sec:RBM}, but presents advantages over the latter since \ac{AE}s can learn more complex data representations.
Furthermore, while the \ac{RBM} is a generative model \citep{hinton2006fast}, the basic \ac{AE} architecture is a feed-forward neural network with the sole purpose of first reducing the dimensionality of the data and then using this to reconstruct the original data. 

The typical structure of an \ac{AE} is defined by two parts. 
The first part, called the encoder $\mathscr{E}_\omega$, maps the input feature space $x$ to a lower-dimensional \textit{latent} space $z$; \mbox{in contrast}, the second part pertains to the decoder $\mathscr{D}_\rho$ which attempts to reconstruct $x$ from $z$, producing $\hat x$.
The optimisation objective of the \ac{AE} is to minimise the difference between the input $x$ and the reconstructed $\hat x$. 
This difference is quantified through various functions, so called loss~functions.
There~exist many types of \ac{AE}s, all based on extensions to the fundamental architecture described until this point.
These extensions can occur in the \ac{AE}'s loss function, at the level of its architecture \citep{Srikumar:2021yzo}, by imposing a structure on its latent space \citep{Kingma:2013hel, joo2020dirichlet}, or combinations thereof \citep{pan2018adversarially, patrini2020sinkhorn}.
For an extensive review of \ac{AE} architectures, see \citet{AEreview}.
In this appendix, a few promising \ac{AE} architectures are identified and presented.

\subsection{Vanilla Autoencoder}
\label{sec:vanillaAE}

\begin{figure*}[t]
    \centering
    \includegraphics[width=0.9\textwidth]{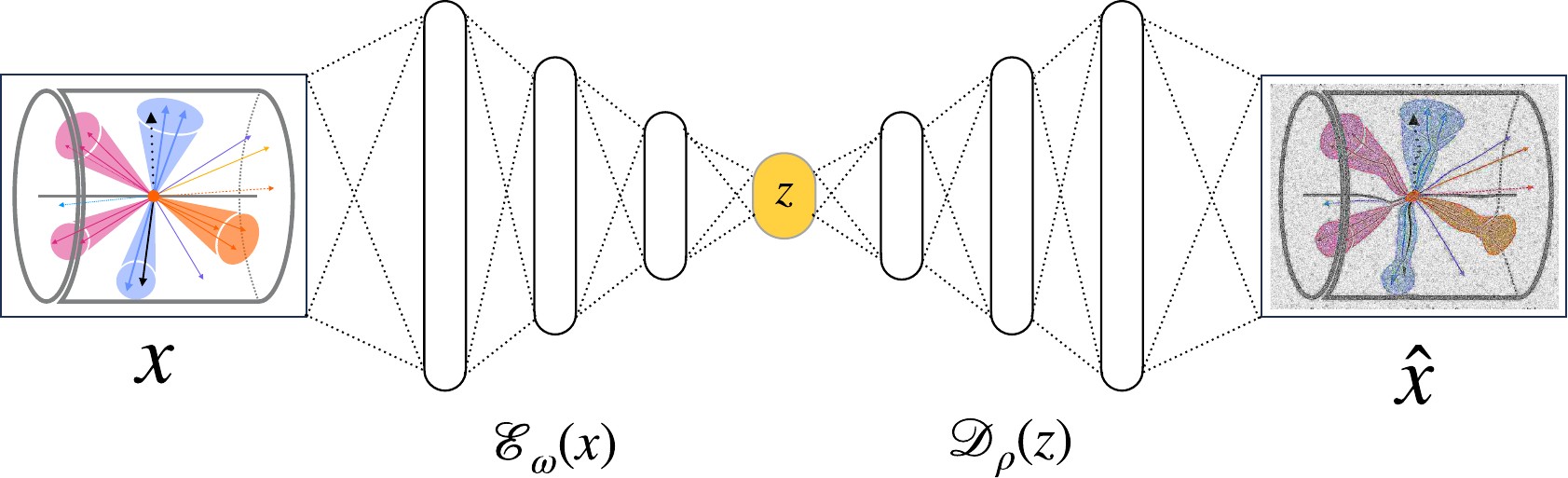}
    \caption{
    Schematic of a conventional autoencoder architecture. 
    The input data $x$ is given to an encoder neural network $\mathscr{E}_\omega$, here depicted with three fully connected layers.
    The neural network processes the data and generates a lower dimensional representation $z$, called the latent space.
    The decoder neural network $\mathscr{D}_\rho$ depicted still with three fully connected layers processes $z$ to generate a reconstruction of the original data $\hat x$.
    The task of this network is to minimise the difference between $x$ and $\hat x$
    }
    \label{fig:vanillaAE}
\end{figure*}

The architecture of the vanilla AE is shown in Fig.~\ref{fig:vanillaAE}.
The encoder $\mathscr{E}_\omega$, defined by trainable parameters $\omega$, maps the $67$ input features to the $16$ dimensional latent space $z$.
The specific dimensionality of $16$ is chosen to be suitable for further processing by the \ac{QML} classifier presented in Sec.~\ref{sec:qsvm}.
The decoder $\mathscr{D}_\rho$, defined by trainable parameters $\rho$, receives $z$ and produces $\hat x \in \mathbf{\hat X}$, a reconstruction of the original input $x\in \mathbf{X}$ that minimises
\begin{equation}
\label{eq:vanillaloss}
\mathcal{L}_\mathrm{MSE} = (x - \hat x)^2.
\end{equation}
The Mean Squared Error (MSE) is then backpropagated~\citep{rumelhart1986learning, damadi2023backpropagation} through both the decoder $\mathscr{D}_\rho$ and the encoder~$\mathscr{E}_\omega$. 
Thus, the structure of the latent space is optimised to facilitate a more faithful input reconstruction by the decoder. 
The conjecture implied by the \ac{AE} is that latent spaces $z$ that better capture the essential information in the original data produce better reconstructions under the action of $\mathscr{D}_\rho$. The hyperparameters of the implemented \ac{AE} architectures are optimised, within computational constraints, such that the minimum achievable $\mathcal{L}_\mathrm{MSE}$ is obtained.
The~resulting architecture is defined by 6 layers for the encoder with widths $\{64, 52, 44, 32, 24, 16\}$ and 6 layers for the decoder with widths $\{24, 32, 44, 52, 64, 67\}$.
The~best batch size and learning rate for the training procedure are found to be 128 and 0.0012, respectively. 

The latent space of a vanilla autoencoder does not follow any particular distribution. 
The extent to which the features of an autoencoder latent space follow a statistical distribution is called the \textit{regularity} of the AE's latent~space. 
The regularity of the vanilla \ac{AE} depends on the input features, the latent space dimension, and the used encoder. 
\mbox{Ultimately}~though, the vanilla encoder will organise the latent space only in such a way that facilitates the reconstruction task and consequently minimises the \ac{MSE} loss from Eq.~\ref{eq:vanillaloss}.

\subsection{Variational Autoencoder}
\label{sec:variationalAE}

\begin{figure*}
    \centering
    \includegraphics[width=0.9\textwidth]{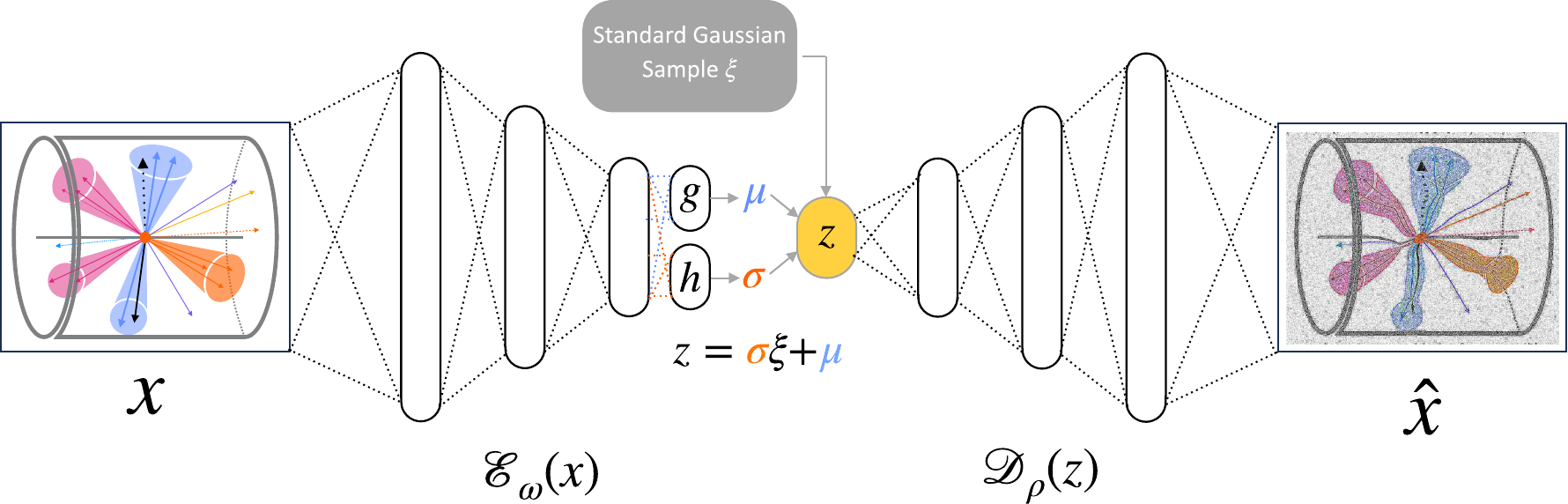}
    \caption{
    Schematic of the variational autoencoder.
    This model functions under the same principles as the vanilla \ac{AE} depcited in Fig.~\ref{fig:vanillaAE}.
    However, the vanilla encoder $\mathscr{E}_\omega$ is connected at the end with two additional layers $g$ and $h$, that generate the mean $\mu$ and standard deviation $\sigma$ of a reference standard Gaussian distribution, respectively.
    The latent space of the variational autoencoder is generated by using these values as shown in the figure.
    Once $z$ is obtained, the decoder $\mathscr{D}_\rho$ proceeds to reconstruct the original data from it.
    Again, the learning goal of this model is to minimise the difference between $x$ and $\hat x$, but also to produce a $z$ that mimics samples from a standard Gaussian distribution $\xi$
    }
    \label{fig:variationalAE}
\end{figure*}

A \ac{VAE} \citep{Kingma:2013hel} is a type of \ac{AE} whose training is regularised to avoid overfitting the data. 
Similar to the vanilla \ac{AE}, the \ac{VAE} is composed of an encoder and a decoder that are trained together to minimise the difference between the reconstructed data and the original features.
However, the \ac{VAE} architecture diverges from the vanilla \ac{AE} architecture in the encoder: to introduce regularisation, each input sample is encoded as a distribution over the latent space. Thus, the \ac{VAE} is called a probabilistic model.

Define $x\in \mathbf{X}$ as the variable that represents the data. 
Then, assume $x$ is generated from a latent variable $z$, where $z$ is not directly observed. 
Accordingly, each data point is generated by first sampling a latent representation $z$ from its prior distribution $p(z)$ and then sampling $x$ from the conditional likelihood distribution $p(x|z)$. 

In this framework, the probabilistic counterpart of the decoder is $\mathscr{D}_\rho \equiv p(x|z)$.
The definition of the probabilistic encoder follows trivially: $\mathscr{E}_\omega \equiv p(z|x)$ returns the distribution of the encoded variable given the decoded one. 
Moreover, latent space regularisation arises naturally: the encoded values $z$ are assumed to follow the prior distribution~$p(z)$. 
For simplicity and since this is the kind of model that is ultimately implemented, assume the following equalities:
\begin{align}
    \label{eq:vaedef}
    p(z) &= \mathcal{N}(0, I) \\
    p(x|z) &= \mathcal{N}(f(z), cI)
\end{align}
where $\mathcal{N}(0, I)$ is the standard normal distribution, $\mathcal{N}(f(z), cI)$ is a Gaussian with mean $f(z)$ and standard deviation $cI$, $I$ is the identity matrix, $c$ is a constant, and $f(z)$ is a deterministic function of the random variable~$z$. 
Furthermore, assume for the moment that $f$ is well defined and fixed. 
Since $p(z)$ and $p(x|z)$ are now defined, Bayes' theorem can in principle be used to get $p(z|x)$ as well; but the computation is intractable~\citep{Kingma:2013hel}. 
Thus, determining $p(z|x)$ necessitates the use of approximation techniques such as variational inference.

The main idea of variational inference is to define a parameterised family of distributions and to search within it for the best approximation of the target distribution, i.e., the distribution to be determined. 
The best approximation is the element of the aforementioned family of distributions that minimises a pre-defined function. 
This function, in principle, measures the error between the trial and the target.
\mbox{Furthermore}, one must also be able to minimise this function through stochastic descent. 
The~function that is most commonly employed for this task is the \ac{KL} divergence 
\begin{equation}
    \label{eq:kldiv}
    D_\mathrm{KL}[q(x)||p(x)] \equiv \int_{x \in\mathcal{X}} q(x)[\log(q(x)) - \log(p(x))]
\end{equation}
for one data class, where $q$ is the target distribution, $p$ is an approximation of the target, and $\mathcal{X}$ the data \citep{goodfellow2016deep, paisley2012variational}. 
Variational inference is applied to approximate $p(z|x)$ by a Gaussian $q_x(z)\equiv \mathcal{N}(g(x), h(x))$, where $g$ and $h$ belong to the families of functions $G$ and $H$. 
The functions $g$ and $h$ are optimised such that $D_\mathrm{KL}(p(z|x),q_x(z))$ is minimised. 

The function $f$ is hitherto assumed to be fixed; however, in practice, the function $f$ is the decoder and is chosen. 
Let $F$ be the group of functions for which the best approximation of $p(z|x)$, denoted $q^*_x(z)$, is obtained for any $f \in F$. 
The most effective encoding-decoding scheme is desired, which means that the function $f$ maximises the probability that for a given input $x$, there exists $\hat{x} = x$ when $z$ is sampled from $q^*_x(z)$ and $\hat{x}$ from $p(x|z)$. 

The functions $f$, $g$, and $h$ can be expressed as neural networks and thus $F$, $G$, and $H$ are the family of functions defined by the possible network architectures. 
Most often, $g$ and $h$ are not defined by two independent networks. 
They share part of their architecture, i.e., $g(x) \equiv g_2(g_1(x))$, $h(x) \equiv h_2(h_1(x))$, and $h_1(x) = g_1(x)$. 
Since $h(x)$ is the covariance of $q_x(z)$, it needs to be a square~\mbox{matrix}. 
Thus, a final approximation is made and $q_x(z)$ is assumed to be a multidimensional Gaussian distribution with a diagonal covariance. 
\mbox{Subsequently}, $h(x)$ becomes a vector containing the diagonal elements of the covariance matrix. 
The $g$ and $h$ functions defined as such make up the encoder $\mathscr{E}_\omega$. 
Furthermore, $p(x|z)$ is defined in Eq.~\ref{eq:vaedef} as a Gaussian with a fixed covariance. 
Thus, the function $f(z)$ defining the mean of said Gaussian can be modelled by a neural network $\mathscr{D}_\rho$ equivalent to the vanilla decoder. 

The \ac{VAE} is the merger between the aforementioned encoder and decoder. 
The architecture of a generic \ac{VAE} is shown in Fig.~\ref{fig:variationalAE}. 
The reparametrisation trick is used to sample latent space values from the encoder: if $z$ is a random variable following a Gaussian distribution with mean $g(x)$ and covariance $h(x)$, then it can be expressed by
\begin{equation}
    z=h(x)\xi + g(x)
\end{equation}
where $\xi$ is sampled from a standard normal distribution. 
This trick allows for the backpropagation of the reconstruction loss, in this case \ac{MSE}, through the network and enables the gradient descent algorithm to function. 
The loss of the \ac{VAE} is derived from the considerations presented in the previous paragraph:
\begin{equation}
    \label{eq:vaeloss}
    \mathcal{L}_\mathrm{VAE} = (1-\alpha)\mathcal{L}_\mathrm{MSE} + \alpha D_\mathrm{KL}(\mathcal{N}(\mu, \sigma), \mathcal{N}(0, I))
\end{equation}
where $\mu$ and $\sigma$ are the outputs of the two branches of the encoder, as depicted in Fig.~\ref{fig:variationalAE}. 
The hyperparameter $\alpha$ is introduced to control how much each loss contributes to the learning.

The hyperparameters of the implemented \ac{VAE} are optimised in the same fashion as for the vanilla \ac{AE}; the hyperparameter $\alpha$ in Eq.~\ref{eq:vaeloss} is set to $0.5$ for the duration of this process.
The best \ac{VAE} encoder consists of 4 layers with the number of nodes in each layer being $\{64, 52, 44, 32\}$, while the decoder has 6 layers with nodes $\{24, 32, 44, 52, 64, 67\}$.
The optimal batch size for the training is determined to be $128$, with a corresponding optimal learning rate of $0.001$. 
Then, the weight $\alpha$ is manually tuned as well: $0.0005$ was determined to give the highest unweighted reconstruction. 
Although this value seems small, the unweighted \ac{KL} divergence loss greatly exceeds $\mathcal{L}_\mathrm{MSE}$. 
Thus, by using such a small weight, the effective contribution from the two losses in the total loss function is similar. 

\subsection{Classifier Autoencoder}
\label{sec:classifierAE}

Implementing the \ac{VAE} architecture led to an improvement in the reconstruction loss, but the distributions of the two data classes overlap in the latent space.
The AUC values of the 16 latent space distributions for both the vanilla \ac{AE} and the variational \ac{AE} are around $0.5 \pm 0.05$, c.f., Sec.\,\ref{sec:results}. 
This similarity between background and signal hinders the performance of any \ac{QML} classifier that takes the \ac{VAE} or the vanilla latent spaces as input.

Hence, a classifier network is attached to the latent space of the vanilla \ac{AE}. 
The new architecture is similar to the one shown in Fig.~\ref{fig:vanillaAE}, but with another neural network attached to the latent space that performs the classification task at the same time as the dimensionality reduction. 
The loss function of this model is defined as
\begin{equation}
    \label{eq:classloss}
    \mathcal{L}_\mathrm{CAE} = (1-\alpha)\mathcal{L}_\mathrm{MSE} + \alpha \mathcal{L}_\mathrm{BCE}
\end{equation}
where $\mathcal{L}_\mathrm{BCE}$ is the \ac{BCE} \citep{goodfellow2016deep} of the classifier and $\alpha$ is the weight hyperparameter, similar to Eq.~\ref{eq:vaeloss}. 
The classifier starts from the $16$ features in the latent space and outputs a single number: the probability that a sample is either signal or background. 

When training this \ac{AE} model with the loss function from Eq.~\ref{eq:classloss}, the encoder learns to produce a latent space that enhances the performance of the classifier.
This encourages a separation between the two classes in the latent space.
The number of layers and the nodes of each layer are kept the same as for the vanilla \ac{AE}.
The classifier attached to the latent space consists of 6 layers with nodes $\{128, 64, 32, 16, 8, 1\}$.
The batch size and learning rate of this model are optimised as in previous sections while setting $\alpha=0.5$. 
The best performing model that we found is trained with a learning rate of $0.001$ and a batch size of $128$ events. 
Further, the weight $\alpha$ is optimised for two different objectives: lowest unweighted $\mathcal{L}_\mathrm{MSE}$ and lowest unweighted $\mathcal{L}_\mathrm{BCE}$, obtaining $\alpha=3\times 10^{-5}$ and $\alpha = 0.6$ respectively.

The obtained models slightly overfit and thus early stopping with a patience of 10 epochs is added to prevent it.
Studying the latent space produced by this classifier \ac{AE} model, we observe that the architecture with $\alpha = 0.6$ tries to encode signal and background at the opposite ends of the latent variable range while its counterpart with $\alpha = 3\times 10^{-5}$ does not display this behaviour, due to the classification branch being heavily suppressed during the training stage.
The overall discriminative power of the latent space produced by the $\alpha = 0.6$ encoder model is significantly better when compared with the alternative.
Finally, the classification performance of the $\alpha=0.6$ model is better than the $\alpha = 1$ model (a pure, albeit unconventional NN classifier) by 0.02.
Thus, it is observed that for the classifier \ac{AE}, trained on the studied \tth data set, the decoder branch and the classifier branch display a ``one-way cooperation''. 
The classification task is improved when the decoder loss $L_\mathrm{MSE}$ is considered, while the reconstruction is always worse when $L_\mathrm{BCE}$ is added to the overall loss function.

\subsection{Sinkhorn Autoencoder}
\label{sec:sinkhornAE}

The architecture of the Sinkhorn AE is illustrated in Fig.~\ref{fig:sinkclassAE}, without the classifier attached to the latent space. 
This~model is implemented from the standard Sinkhorn AE \citep{patrini2020sinkhorn}. 
The optimisation objective of the Sinkhorn \ac{AE} is
\begin{equation}
    \label{eq:sinkloss}
    \mathcal{L}_\mathrm{SAE} = \alpha \mathcal{L}_\mathrm{SH} + (1-\alpha)\mathcal{L}_\mathrm{MSE} 
\end{equation}
where $\mathcal{L}_\mathrm{SH}$ is an approximation of the Wasserstein distance between the output of the conditional noise generator and the encoded input. 
This new metric is added to stabilise the training of generative adversarial neural networks \citep{arjovsky2017wasserstein}. 
Since then, the Wasserstein \ac{AE} \citep{tolstikhin2017wasserstein} appeared, where $\mathcal{L}_\mathrm{SH}$ is used to regularise the latent space instead of $D_\mathrm{KL}$. 

To understand the Wasserstein distance, consider a data distribution $q$ and a model distribution $p$ can be matched by minimising the \textit{optimal transport}~cost between them. 
Two~distributions can be related by constructing a joint distribution $\Gamma$, also called the transport plan since it moves data samples around to reshape $p$ into $q$. 
The~transport plan is not unique and can be chosen depending on the scope of the problem. 
Furthermore, the transport of each sample inside a distribution has an associated cost: the farther the sample is moved, the larger the cost.

The Wasserstein distance or divergence is defined as
\begin{equation}
    W_c(q,p) = \inf_{\Gamma \in \Pi(q,p)} \iint c(x,y)\Gamma(x,y)dxdy
\end{equation}
where $\Pi(q,p)$ is the complete set of transport plans between $q$ and~$p$. 
Thus, $W_c$ is the cost of reshaping $p$ into $q$ under the cheapest (optimal) transport plan. 
The actual calculation of $W_c$ and more complex versions thereof is computationally expensive. 
The Sinkhorn algorithm is used to approximate $W_c$ in an efficient way; for details of how this approximation is calculated, see~\citet{patrini2020sinkhorn}. 
The result is the so called Sinkhorn loss $\mathcal{L}_\mathrm{SH}$. 
In practice, the efficient implementation of the Sinkhorn algorithm with GPU acceleration from the GeomLoss package \citep{feydy2019interpolating} is used to approximate $W_c$. 
The gradient obtained from $\mathcal{L}_\mathrm{SH}$ is employed to simultaneously train the encoder and the noise generator from Fig.~\ref{fig:sinkclassAE}, bringing them towards convergence into a similar distribution in the latent space. 

The learning rate and batch size of the implemented end-to-end Sinkhorn AE are optimised while $\alpha=0.5$ is fixed. 
The values obtained thus are $0.001$ and $128$, respectively. 
The encoder and the decoder have the same structure as the vanilla \ac{AE}.
The standard gaussian sample is passed through a network with two layers and with nodes $\{64, 128\}$.
The input truth labels are passed through a one layer network with 64 nodes.
Subsequently, the outputs of the last two components are processed by a three layer network with nodes $\{256, 192, 16\}$.
These are then fixed and $\alpha$ is in turn optimised. 
The model with the lowest unweighted $\mathcal{L}_\mathrm{MSE}$ value has $\alpha=0.06$. 
Thus, a latent space that is more regular than the vanilla \ac{AE} but less constrained than the \ac{VAE} one is obtained.

\end{document}